\documentclass[english,12pt]{iopart}
\usepackage[T1]{fontenc}
\usepackage[latin9]{inputenc}
\usepackage{float}
\usepackage{units}
\usepackage{amstext}
\usepackage{graphicx}
\usepackage{amssymb}
\usepackage[numbers,sort&compress]{natbib}

\makeatletter

\DeclareRobustCommand{\greektext}{%
  \fontencoding{LGR}\selectfont\def\encodingdefault{LGR}}
\DeclareRobustCommand{\textgreek}[1]{\leavevmode{\greektext #1}}
\DeclareFontEncoding{LGR}{}{}
\DeclareTextSymbol{\~}{LGR}{126}
\newcommand{\lyxmathsym}[1]{\ifmmode\begingroup\def\b@ld{bold}
  \text{\ifx\math@version\b@ld\bfseries\fi#1}\endgroup\else#1\fi}

\usepackage{iopams}
\usepackage{setstack}

\makeatother

\usepackage{babel}

\begin{document}
\title[THz-driven quantum wells]{THz-driven quantum wells: Coulomb interactions and Stark shifts in the ultrastrong coupling regime}

\author{Benjamin Zaks$^1$$^,$$^2$, Dominik Stehr$^1$$^,$$^3$, Tuan-Anh
Truong$^4$, Pierre M. Petroff$^4$, Stephen Hughes$^5$, Mark S.
Sherwin$^1$$^,$$^2$}

\address{$^1$ Institute for Terahertz Science and Technology, University
of California at Santa Barbara, Santa Barbara CA 93106}

\address{$^2$ Physics Department, University of California at Santa Barbara,
Santa Barbara CA 93106}

\address{$^3$ Institute for Ion Beam Physics and Materials Research, Helmholtz-Zentrum
Dresden-Rossendorf, P.O. Box 510119, 01314 Dresden, Germany}

\address{$^4$ Materials Department, University of California at Santa Barbara,
Santa Barbara CA 93106}

\address{$^5$ Department of Physics, Queen's University, Kingston, Ontario,
K7L 3N6 Canada}

\ead{bzaks@physics.ucsb.edu}
\begin{abstract}
We investigate the near infrared interband absorption of semiconductor
quantum wells driven by intense terahertz radiation in the regime
of ultrastrong coupling, where the Rabi frequency is a significant
fraction of the frequency of the strongly driven transition. With
the driving frequency tuned just below the lowest frequency transition
between valence subbands, a particularly interesting phenomenon is
observed. As the THz power increases, a new peak emerges above the
frequency of the undriven exciton peak which grows and eventually
becomes the larger of the two. This reversal of relative peak intensity
is inconsistent with the Autler-Townes effect in a three-state system
while within the rotating wave approximation (RWA). In the samples
investigated, the Bloch-Siegert shift (associated with abandoning
the RWA), exciton binding energy, the Rabi energy, and non-resonant
AC Stark effects are all of comparable magnitude. Solution of a semiconductor
Bloch model with one conduction and multiple valence subbands indicates
that the AC Stark effect is predominantly responsible for the observed
phenomenon.
\end{abstract}

\pacs{78.67.De, 42.50.Hz, 71.35.-y, 78.40.Fy}

\submitto{\NJP}

\maketitle

\section{Introduction}

Strong interactions between light and matter are historically and
currently of great interest in condensed matter, atomic and molecular
physics as well as in quantum information science. In the strong coupling
regime, effects of the light-matter interaction are manifested as
coherent Rabi oscillations between a pair of quantum states \citep{Rabi1937,Cundiff1994},
vacuum Rabi splitting of two-state quantum systems in a cavity \citep{Reithmaier2004,Yoshie2004,Thompson1992},
and the Autler-Townes effect in three-state quantum systems \citep{Autler1955,Dion1976,Kamada2001,Xu2007,Carter2005,Dynes2005,Wagner2010,Baur2009,Sillanpaa2009}.
As many of these systems have proposed uses in the coherent control
of states and in quantum information processing \citep{Mabuchi2002},
further studies of light-matter coupling pave the way for future technologies.
Recently, cavity-coupled semiconductor \citep{Guenter2009,Todorov2010}
and superconductor \citep{Niemczyk2010} systems have achieved the
ultrastrong coupling regime where the photon exchange between the
atom and optical field occurs at a rate (given by the Rabi frequency,
$\Omega_{\textrm{Rabi}}$) comparable to the frequency of the photon.
Though ultrastrong coupling has been difficult to achieve in the past,
recent experiments open the door to investigate effects such as the
Bloch-Siegert shift \citep{Bloch1940,Arimondo1973,Tuorila2010,Forn-D'iaz2010}
and carrier-wave Rabi flopping \citep{Hughes1998,Mucke2001} which
are expected to become observable in this regime. With an intense
THz field incident upon a semiconductor quantum well, we are able
to vary the light-matter coupling strength and investigate the ultrastrong
coupling limit without a cavity.

With intense electromagnetic (EM) fields tuned resonantly to a two-level
quantum transition, Autler and Townes found that absorption of a weak
probe between one of the strongly driven levels and a third state
split into two symmetric peaks. The separation between the peaks can
act as a direct measure of the strength of the coupling between the
atom and the field and has a magnitude of the Rabi energy, $\hbar\Omega_{\textrm{Rabi}}=2\mu E$,
where $\mu$ is the dipole moment between the strongly driven states
and $E$ is the amplitude of the EM field. Autler-Townes splitting
has been studied in atomic and molecular systems for decades \citep{Autler1955,Dion1976},
but has only been observed in low-dimensional semiconductors \citep{Kamada2001,Xu2007,Carter2005,Dynes2005,Danielson2007,Jameson2009,Wagner2010}
and in superconducting quantum systems \citep{Baur2009,Sillanpaa2009}
in the last ten years. In semiconductors, effects such as Coulomb
interactions between electrons and holes enrich light-matter interactions
\citep{Kira2006} with many-body effects.

Semiconductor heterostructures are particularly attractive for investigating
intense light-matter interactions since - unlike atoms - the quantum
energy levels and effective masses can be engineered at the growth
stage, while the dimensionality of the system can be uniquely controlled
by quantum confinement. The Autler-Townes effect in semiconductors
has been investigated in both quantum well \citep{Carter2005,Dynes2005}
and quantum dot systems \citep{Kamada2001,Xu2007,Boyle2009}. At near
infrared frequencies, narrow excitonic linewidths in quantum dots
led to the observation of small values of Rabi splitting ($\sim\unit[1.5]{\mu eV}$)
at low powers ($\sim\unit[1]{W/cm^{2}}$) \citep{Xu2007}. More intense
sources ($\sim\unit{MW/cm^{2}}$) were required for observation of
the Autler-Townes effect in quantum wells (QWs). With these intense
fields ($\sim\unit[10]{kV/cm}$), Rabi energies of a few meV were
seen in both intersubband absorption spectra of doped QWs driven at
mid infrared (IR) frequencies \citep{Dynes2005} and in the interband
absorption of undoped QWs driven at THz frequencies \citep{Carter2005,Wagner2010}.
Rabi energies which are a significant fraction of the transition energy
were observed by Carter \etal\citep{Carter2005} and indicate that
THz-driven quantum wells are a well-controlled system in which to
study ultrastrong light-matter coupling.

\begin{figure}[t]
\begin{centering}
\includegraphics{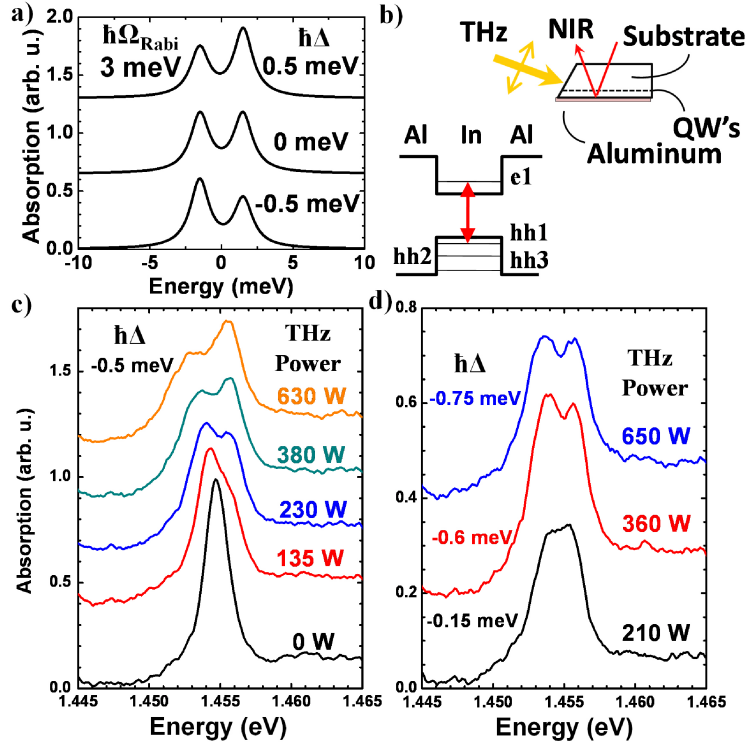}
\par\end{centering}

\caption{a) Absorption spectra from a strongly driven two-level system that
is weakly probed from a third state calculated in the rotating wave
approximation (RWA). The transition energy $\hbar\omega_{\textrm{2-1}}$
is 6 meV. Spectra are shown for three values of the detuning $\Delta=\omega_{\textrm{THz}}\lyxmathsym{\textendash}\omega_{\textrm{2-1}}$.
b) A schematic showing the geometry of the sample. Also shown is a
simple band diagram with the energy levels relevant to the experiment.
The red arrow indicates the optically active transition. c) Experimentally
observed absorption spectra of the 22 nm quantum wells for increasing
THz powers. The powers quoted in the figures have not been adjusted
to account for reflection or absorption from the sample facet or the
cryostat window, but provide an accurate measure of the relative intensity
at the wells. The energy of the driving field $\hbar\omega_{\textrm{THz}}$
is tuned 0.5 meV below the 6 meV exciton transition $\hbar\omega_{\textrm{X2-X1}}$.
d) Symmetric absorption spectra from 22 nm QWs at a number of values
of negative detuning demonstrating that intense THz fields red shift
$\omega_{\textrm{X2-X1}}^{*}$ into resonance with $\omega_{\textrm{THz}}$.
Resonance was observed at a detuning as large as $\hbar\Delta=\hbar(\omega_{\textrm{THz}}-\omega_{\textrm{X2-X1}})=\unit[-0.75]{meV}$,
corresponding to a shift, $\frac{(\omega_{\textrm{X2-X1}}-\omega_{\textrm{X2-X1}}^{*})}{\omega_{\textrm{X2-X1}}}$,
greater than 12\%.\label{Experiment}}

\end{figure}

The Autler-Townes effect in a three-level system can be solved analytically
within the commonly-used rotating wave approximation (RWA) \citep{Rabi1954}.
In this model, the splitting in the absorption spectrum is symmetric
when the driving frequency $\omega_{\textrm{THz}}$ is resonant with
the transition frequency $\omega_{\textrm{2-1}}$ (\fref{Experiment}(a)).
With a non-resonant driving frequency, the symmetry of the absorption
spectrum is broken and the relative peak amplitudes are determined
by the detuning, $\Delta=\omega_{\textrm{THz}}-\omega_{\textrm{2-1}}$.
For the level probed in the following experiments, a positive (negative)
detuning results in a larger (smaller) peak at higher energy. In the
RWA model, $\omega_{2-1}$ is independent of the strength of the driving
field.

The RWA breaks down in the limit of ultrastrong coupling. The
effect of abandoning the RWA in a two-level system was first investigated
by Bloch and Siegert \citep{Bloch1940}. With increasing resonant EM
field amplitude, they found a blue shift of the effective transition
frequency, $\omega_{\textrm{2-1}}^{*}\left(\mbox{\ensuremath{\Omega}}_{\textrm{Rabi}}\right)=\omega_{2-1}\left(1+\frac{\Omega_{\textrm{Rabi}}^{2}}{4\omega_{\textrm{2-1}}^{2}}\right)$.
This effect will become observable in the ultrastrong coupling regime,
when $\Omega_{Rabi}$ is a significant fraction of $\omega_{2-1}$.
In the three-level system under investigation, as the EM field amplitude
is increased the effective transition energy will increase and the
system will become negatively detuned. As the magnitude of the negative
detuning increases, the amplitude of the higher energy Autler-Townes
peak will $\mathit{decrease}$ as a result of the Bloch-Siegert blue-shift
\citep{Zhang2004}.

In this work a systematic investigation of Autler-Townes splitting
for several quantum well systems is described. Though the system is
in the ultrastrong coupling regime ($\nicefrac{\Omega_{Rabi}}{\omega_{2-1}}\sim.5$)
where the Bloch-Siegert effect is non-negligible, it is observed experimentally
that the amplitude of the peak at higher energy $\mathit{increases}$
with increasing terahertz intensity, indicating a red shift of $\omega_{\textrm{2-1}}^{*}$.
At first sight this $\mathit{reverse\, Bloch-Siegert\, shift}$ is
a complete surprise. By solving a full four-subband semiconductor
Bloch model outside the RWA, theoretical results that are in good
agreement with experiments on two different QWs are achieved. These
calculations show that effects due to non-RWA terms in the Hamiltonian
and effects due to higher lying levels both persist in these samples.
An AC Stark shift is predominantly responsible for the observed phenomena.

\section{InGaAs quantum well absorption}

\subsection{Experimental observations}

The experimental design was similar to that of Carter \etal \citep{Carter2005},
where a strong THz field polarized in the growth direction of the
undoped InGaAs/AlGaAs QWs dramatically modified the NIR interband
absorption spectrum via the Autler-Townes effect. Employing thicker
QWs with lower transition energies than in previous experiments allowed
investigation further into a regime where the RWA breaks down. The
two samples investigated in these experiments were grown by molecular
beam epitaxy with 10 repetitions of In$_{0.06}$Ga$_{0.94}$As quantum
wells of 18 and 22 nm widths, respectively, separated by 30 nm Al$_{0.3}$Ga$_{0.7}$As
barriers. The samples were then capped by 50 nm of GaAs. Strain in
the samples shifted the valence band energies and isolated the lowest
heavy hole states from the light holes. Compared to the heavy holes,
the electrons have a low effective mass, leading to much a larger
energy separation in the conduction band than in the valence band.
Thus the system could be modeled as a single electron level interacting
with a few heavy hole levels (\fref{Experiment}(b)). The facet
was polished at 10 degrees and a $\unit[\sim250]{nm}$ thick Aluminum
layer was deposited to optimize the coupling of the THz E-field to
the QWs. The Aluminum layer also allowed the sample to be studied
in double pass geometry by providing a reflective layer for the NIR
probe. The THz radiation was generated by the UCSB Far-Infrared Free-Electron
Laser (FEL) \citep{Ramian1992}, a narrowband (<1 GHz) source capable
of producing pulses with kilowatt power and few-$\unit{\mu s}$ duration
at the frequencies of interest in these experiments, 1.1 THz to 2.5
THz. A small percentage of the beam was sent to a pyroelectric detector
to monitor the THz power while the power at the sample was controlled
by a pair of wire grid polarizers. The diameter of the focused THz
beam was $\sim\unit[1]{mm}$, providing a THz intensity of $\sim\unit[100]{kW/cm^{2}}$
in the sample.

In the absence of a THz driving field, the single strong NIR absorption
line was assigned to the exciton X1 consisting of an electron from
the lowest conduction subband (e1) and a hole from the highest valence
subband (hh1). The X2 (e1-hh2) exciton was not observed because initial
and final states have the same parity. The X3 (e1-hh3) exciton was
not observed because of small transition matrix elements. The observed
plateau towards higher energy arises from the continuum absorption.
Experiments showed that the most interesting behavior was observed
when the driving frequency $\omega_{\textrm{THz}}$ was tuned slightly
below the resonant frequency of the exciton transition $\omega_{\textrm{X2-X1}}$
\citep{Sherwin2007} (\fref{Experiment}(c)). At low intensities,
the peak at higher energy was smaller, the expected behavior for negative
detuning (\fref{Experiment}(a), lower trace). As the THz intensity
was increased, the amplitude of the higher energy peak increased.
At THz powers above 380 W, the peak at higher energy was the larger
peak, the expected behavior for positive detuning (\fref{Experiment}(a),
top trace). The intensity dependent change of the relative peak amplitudes
was interpreted as a red shift of the effective exciton transition
frequency $\omega_{\textrm{X2-X1}}^{*}$. Experiments were performed
on this sample at a number of THz frequencies below resonance (\fref{Experiment}(d)) and it was found that larger initial values
of negative detuning required more intense THz fields to create a
symmetric splitting, further verification that strong THz fields produced
a red shift of $\omega_{\textrm{X2-X1}}^{*}$.

\subsection{Theoretical description}

To compute the THz-driven spectra, a numerical solution of the semiconductor
Bloch equations was performed in the presence of both a weak optical
probe field and a strong THz driving field. The approach was similar
to that of Liu and Ning \citep{Liu2000}, who included a total of three
subbands in their calculation and made the RWA. Previous theoretical
studies of THz-driven quantum wells have been performed without making
the RWA, but have not focused on the Bloch-Siegert effect \citep{Maslov2000,Maslov2002}.
Calculations presented here included a single conduction subband,
either two or three valence subbands, and the continuum states of
a symmetric quantum well. An efficient numerical solution of the semiconductor
Bloch equations was carried out in real space \citep{Hughes2004} without
application of the RWA to the THz fields. Using an infinite QW model,
energy levels were calculated and exciton effects were introduced
by a self consistent determination of the binding energies and Coulomb
matrix elements. Calculations neglected band mixing and took parameters
typical of an InGaAs/AlGaAs system (Appendix D). All simulations used
an exciton dephasing rate for the off-diagonal transitions that resulted
in a homogeneous broadening rate (FWHM) $\gamma_{\textrm{hom}}=\unit[1.4]{meV}$.
Simulations shown in \fref{Theory and Experiment} also include
an inhomogeneous broadening $\gamma_{\textrm{inh}}=\unit[0.8]{meV}$
(Appendix A).

Self-consistent determination of the exciton binding energies revealed
that X1 has a greater binding energy than X2. This resulted in the
inter-exciton transition energy $\hbar\omega_{\textrm{X2-X1}}$ being
larger than the bare intersubband transition energy $\hbar\omega_{\textrm{2-1}}$,
an effect also predicted by Liu and Ning \citep{Liu2000}. Thus the
22 nm quantum well, with a calculated hh2-hh1 intersubband transition
energy of $\hbar\omega_{\textrm{2-1}}=\unit[5.3]{meV}$, has an exciton
transition energy of $\hbar\omega_{\textrm{X2-X1}}\approx\unit[6]{meV}$.
This significant difference in the energy spectra exemplifies the
deviation of the excitonic system from the intersubband system and
implies that Coulomb effects are important in observed and calculated
spectra.

\begin{figure}
\begin{centering}
\includegraphics{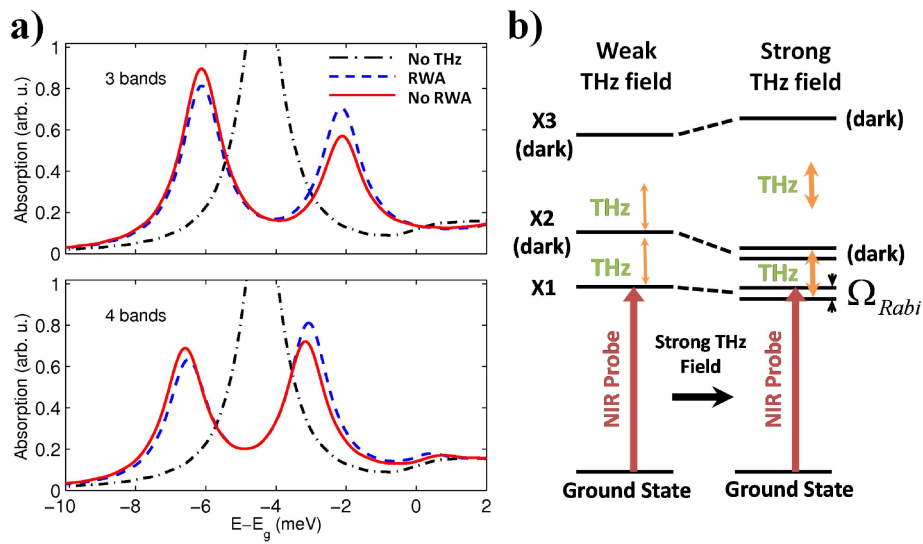}
\par\end{centering}

\caption{a) Simulations of the Autler-Townes splitting of an optically excited
22 nm quantum well with $\hbar\omega_{\textrm{X2-X1}}=\unit[6]{meV}$,
$\hbar\omega_{\textrm{THz}}=\unit[5.5]{meV}$ and $\hbar\Omega_{\textrm{Rabi}}=\unit[4]{meV}$.
The black-chain line shows the absorption spectrum with no THz driving.
The top graph shows simulations calculated with only three subbands
while the bottom graph also includes the fourth subband. The Bloch-Siegert
shift is evidenced by differences in simulations calculated with (blue
dashed) and without (red solid) the RWA. Exciton dephasing rates are
set to produce a homogeneous broadening $\gamma_{\textrm{hom}}$ =
1.4 meV. b) An exciton energy level schematic showing the effect of
a strong, negatively detuned THz field on the states. In the 4 subband
model, a strong Stark-like shift to the X3-X2 transition overcomes
the Bloch-Siegert effect and produces an overall red shift of the
X2-X1 transition.\label{Theory}}

\end{figure}

Simulations of a 22 nm quantum well with $\omega_{\textrm{THz}}$
tuned below $\omega_{\textrm{X2-X1}}$ are shown in \fref{Theory}(a).
Three-subband simulations of the Autler-Townes splitting with (blue
dashed) and without (red solid) the RWA are shown (\fref{Theory}(a),
top graph); the chain curve shows the absorption in the zero THz limit.
Though both simulations exhibit a negative detuning, the difference
between the heights of the two peaks is larger in the non-RWA spectrum,
showing that it is further from resonance than the RWA spectrum. This
difference between the effective detuning $\omega_{\textrm{THz}}-\omega_{\textrm{X2-X1}}^{*}$
of the two spectra is evidence of a Bloch-Siegert effect which blue
shifts $\omega_{\textrm{X2-X1}}^{*}$ in the non-RWA model. By itself,
this is an interesting result: the Bloch-Siegert effect, calculated
for a simple three-state system, persists in the 3-subband SBE model
with Coulomb interactions. However, the persisting negative detuning
of the system at this intense value of the Rabi energy, even without
the RWA, shows the absence of a red shift and demonstrates that the
three-subband model is not consistent with our experimental findings.

Finally, the third valence subband (hh3) was introduced which allowed
coupling between excitons X2 and X3. Consistent with the infinite
well approximation, the subband hh3 was separated from hh2 in energy
by 9.5 meV ($\frac{5}{3}\hbar\omega_{\textrm{2-1}}$), more than 4
meV off resonance from the driving frequency. As is seen in the three
subband model, simulations (\fref{Theory}(a), bottom graph)
indicate a Bloch-Siegert shift of the non-RWA spectra. With an intense
THz field applied at a frequency which is $below$ that of the exciton
transition $\omega_{\textrm{X2-X1}}$, the relative amplitudes of
the two Autler-Townes peaks indicate that the system is being driven
$above$ resonance. This behavior is consistent with our experimental
observations and implies that the strong THz field induces a redshift
of the effective transition energy $\omega_{\textrm{X2-X1}}^{*}$.
This red shift is interpreted as a Stark-like shift to the off resonant
X3-X2 transition which overcame the expected Bloch-Siegert effect,
diagrammed by an energy level schematic in \fref{Theory}(b).
In summary, the combination of non-RWA effects and a fourth subband
in the SBE model predicted a red shift of $\omega_{\textrm{X2-X1}}^{*}$
which was in good agreement with the experimental data.

\subsection{Results}

\begin{figure}[H]
\begin{centering}
\includegraphics{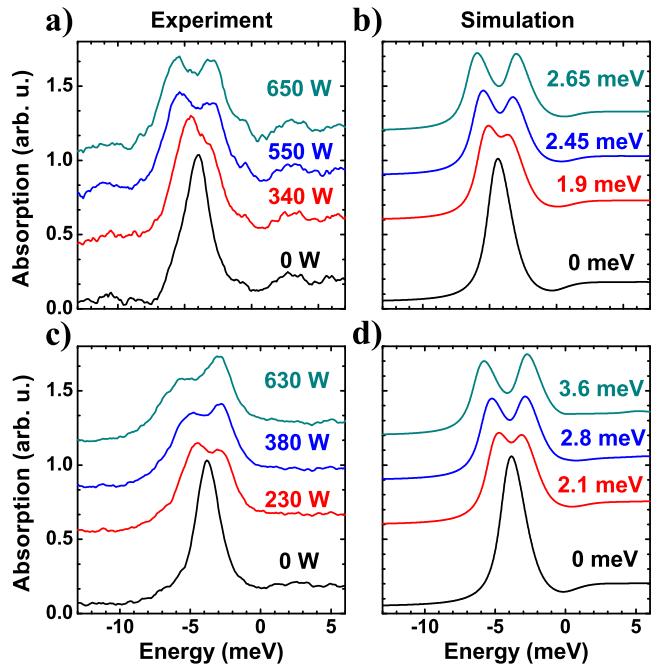}
\par\end{centering}

\caption{Experimental and calculated interband absorption spectra from two
QWs at multiple powers with FEL detuned slightly below resonance.
Experimental data from an 18 nm QW driven 0.4 meV below the 10 meV
excitonic transition $\hbar\lyxmathsym{\textgreek{w}}_{\textrm{X2-X1}}$
is shown in a) with curves labeled by THz power. Simulations are plotted
in b) with curves labeled by the Rabi energy. Experimental results
from the 22 nm QW are shown in c) with corresponding simulations in
d). All simulations use $\lyxmathsym{\textgreek{g}}_{\textrm{hom}}=\unit[1.4]{meV}$
and $\lyxmathsym{\textgreek{g}}_{\textrm{inh}}=\unit[0.8]{meV}$.
\label{Theory and Experiment}}

\end{figure}

Experimental results from 18 and 22 nm wide QWs, both driven slightly
below resonance, are shown in \fref{Theory and Experiment}(a)
and \fref{Theory and Experiment}(c), respectively. When the 18 nm
QW was driven with the THz frequency negatively detuned 0.4 meV below
$\hbar\lyxmathsym{\textgreek{w}}_{\textrm{X2-X1}}=\unit[10]{meV}$,
a symmetric splitting was achieved at 650W. A Rabi energy $\hbar\Omega_{\textrm{Rabi}}=\unit[2.7\pm0.1]{meV}$
was estimated by a fit of two Lorentzians to the absorption spectrum
(Appendix B). Calculations of the 18 nm QW (\fref{Theory and Experiment}(b))
with a symmetric splitting $\hbar\Omega_{\textrm{Rabi}}=\unit[2.65]{meV}$,
a value within the experimental error of the experimentally observed
value, show strong agreement with the experiment. For simulations
at lower powers, the Rabi energy used in the simulations was scaled
to the square root of the experimentally measured power levels (Appendix
B). The resulting absorption curves reproduce the behavior observed
in experiment quite well, namely, the increasing amplitude of the
high-energy peak with increasing THz intensity.

Calculations of the 22 nm QWs driven 0.5 meV below $\hbar\lyxmathsym{\textgreek{w}}_{\textrm{X2-X1}}=\unit[6]{meV}$
clearly show the reversal of relative peak intensities with increasing
Rabi energy that is observed in the experiments (\fref{Theory and Experiment}(d)).
The splitting at 630 W was fit to two Lorentzians and found to be
$\sim3$ meV, so the 3.6 meV simulations are at more intense fields
than in the experiment but still demonstrates the behavior observed
in experiments. The amplitude of the higher energy peak is larger
in the experimental data than in the simulations, so the calculated
red-shift of $\hbar\lyxmathsym{\textgreek{w}}_{\textrm{X2-X1}}^{*}$
was not as large the experimental shift. However, the reversal of
peak intensities, evidence of a redshift of the resonance, is still
clearly replicated at high fields. Further simulations expanding our
model to four valence subbands were performed, but these had qualitatively
similar results to the three subband model. Heating has been thoroughly
investigated in this and previous experiments and is not responsible
for the behavior observed in this experiment (Appendix C). The overall
success of our final model showed that the THz-driven QW system is
modified by competing effects and a proper description must include
a conduction subband, at least three valence subbands and must not
invoke the RWA.

\section{Conclusion}

We have studied two QW samples in the limit of ultrastrong coupling.
In the widest QWs investigated, the Rabi energy approached half of
the energy of the strongly driven transition $\hbar\lyxmathsym{\textgreek{w}}_{\textrm{X2-X1}}^{*}$.
Though a blue-shift of the transition is expected due to non-RWA effects,
what is observed is a red-shift which exceeds 12\% of the transition
energy. Theoretical investigations reveal the interplay between the
Bloch Siegert shift and the AC Stark effect and indicate that the
Stark shift is the dominant effect in these samples. This interplay
can be further investigated with semiconductor QWs grown in different
geometries. When the energy of the second lowest transition $E_{32}\gtrsim E_{21}$,
as is the case in square quantum wells like the ones investigated,
the AC Stark shift is expected to be dominant. If the energy $E_{\textrm{32}}\gg E_{\textrm{21}}$,
as is the case for tunnel-coupled double QWs, we expect the Stark
shift would be negligible and only the Bloch-Siegert shift would be
observed. Similarly, if $E_{32}<E_{\textrm{21}}$, as in triangle
QWs, the AC Stark shifts is a blue shift and the Stark effect is in
the same direction as the Bloch-Siegert effect. Our results, coupled
with the adaptability of the terahertz-driven quantum well system,
implies that this is an ideal system to investigate ultrastrong coupling
on a quantitative level and to controllably explore the regime outside
the three-level model and beyond the RWA.

\addcontentsline{toc}{section}{Acknowledgments}

\ack{}{This research was funded by NSF-DMR grants 0703925 and 1006603, the
Alexander-von-Humboldt Foundation (DS) and the Natural Sciences and
Engineering Research Council of Canada. We thank Prof. David Citrin
for stimulating discussions as well as for introducing SH to this
collaboration, Dr. Alex Maslov for early calculations, Prof. Craig
Pryor for appropriate effective masses, Chris Morris for many critical
readings of this manuscript, and David Enyeart for assistance running
the UCSB FEL.}

\appendix
\setcounter{section}{1}
\section*{Appendices}
\section*{Appendix A. Broadening}
\addcontentsline{toc}{section}{Appendix A. Broadening}

The true linewidth seen in experiments is caused by both homogeneous
and inhomogeneous broadening, so the linewidth alone cannot be used
to estimate the exciton dephasing rate, the source of homogeneous
broadening. However, increasing the exciton dephasing rate broadens
the exciton line and reduces the maximum amplitude of the exciton
absorption while having little effect on the continuum absorption.
Therefore, to obtain an estimate of the dephasing rate, the rate is
adjusted until the relative strengths of the continuum and exciton
absorptions in the simulations fit that of the experimental data when
no THz field applied. This led to an exciton dephasing rate for the
off-diagonal transitions that resulted in a homogeneous broadening
(FWHM) $\gamma_{\textrm{hom}}=1.4\,\textrm{meV}$. Terahertz induced
broadening arose directly from the SBE simulations where $\gamma_{\textrm{hom}}$
was held constant for all values of the Rabi energy. In addition,
it was found that small changes to the homogeneous broadening have
little effect on our predictions.

\begin{figure}[h]
\begin{centering}
\includegraphics{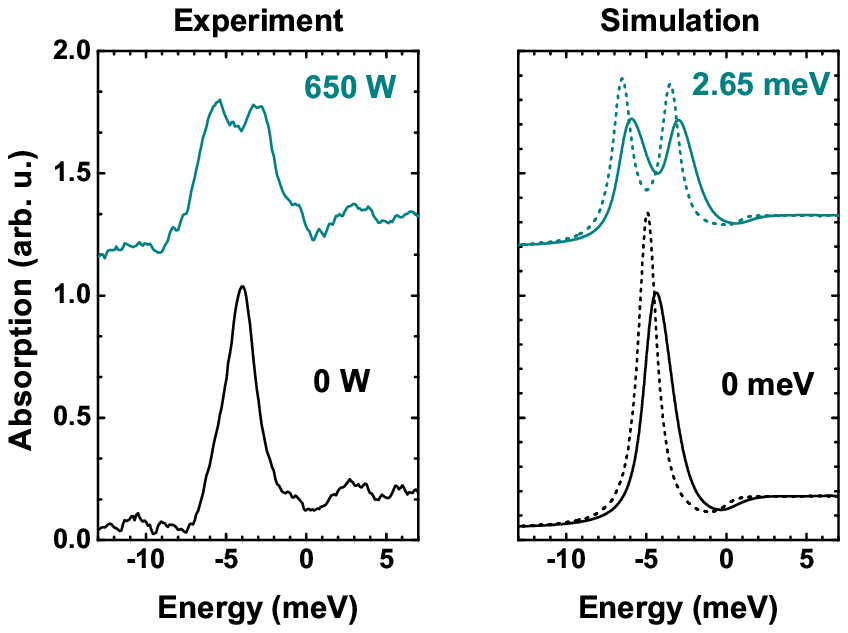}
\par\end{centering}

\caption{Experimental and simulated absorption spectra. The simulated spectra
are shown with (solid lines) and without (dashed lines) inhomogeneous
broadening $\gamma_{\textrm{inh}}=0.8$ meV. Inhomogeneously broadened
spectra are more representative of the experimentally observed absorption.\label{Broadening}}

\end{figure}

Inhomogeneous broadening was qualitatively included by convolving
the SBE solution with a Gaussian function, an effect that produced
spectra more representative of the experimental results. Due to the
continuum absorption at higher energy than the exciton peak, convolving
the simulated spectra with a Gaussian function shifts the exciton
peak towards higher energies. Simulations for both QWs are shown with
(solid lines) and without (dashed lines) an inhomogeneous broadening
$\gamma_{\textrm{inh}}=0.8\,\textrm{meV}$ (\fref{Broadening}).

\section*{Appendix B. Dipole moments, electric field and Rabi energy}
\addcontentsline{toc}{section}{Appendix B. Dipole moments, electric field and Rabi energy}

The dipole moments used in the SBE simulations are calculated directly
from an infinite potential well model. For an intersubband transition
the dipole moment $\mu$ is given by $e\left\langle z\right\rangle $,
where $e$ is the electric charge and $\left\langle z\right\rangle $ is
the matrix element of the position operator in the growth direction.
For the 22 nm QW, the dipole moments of the two lowest transitions
are $z_{\textrm{21}}=3.97$ nm and $z_{\textrm{32}}=4.28$ nm. The
18 nm QW has $z_{\textrm{21}}=3.24$ nm and $z_{\textrm{32}}=3.50$
nm. These dipole moments are very useful in verifying that the electric
fields being used in the SBE calculations are consistent with what
we believe we are applying experimentally. 

It is very difficult to get an accurate estimate of the electric field
at the sample using the power in the THz beam due to the complicated
geometry where the beam strikes the sample facet. However, assuming
the dipole moment is constant, as is done in the simulations, the
Rabi energy is directly proportional to the electric field. We also
stated that the magnitude of the splitting, when the driving beam
is resonant, is given by the Rabi energy. Because we have the ability
to measure the splitting quite accurately by fitting two Lorentzian
peaks to the experimental data, we can obtain an accurate measure
of the electric field from the splitting. 

From the Lorentzian fits, we measure the maximum Rabi splitting in
the 18 nm QW to be 2.7 meV. Using this Rabi energy and the dipole
moment calculated above, we estimate the electric field at the QW
to be $~4.1$ kV/cm. Calculations of the electric field based on the
power and the diameter of the focused THz beam estimate the field
strength to be between 4 and 5 kV/cm, a figure that is in good agreement
with the value estimated from the Rabi splitting. For calculations
of the electric field in the sample, values used are: the power in
the sample $P_{\textrm{in}}\approx\,.6*\textrm{Power\,\ measured ($\sim3$ kW)}$,
beam diameter $d\,\approx\,800\,\mu m$, index of refraction $n=3.3$,
spot area $A=\frac{\pi d^{2}}{4}$, and field strength $F=\sqrt{\frac{2P_{\textrm{in}}}{A\varepsilon_{0}cn}}$.
This implies that it is reasonable to use the Rabi splitting to estimate
the Rabi energy, and to use these values in the simulations when calculating
the corresponding spectra.

To obtain the Rabi energies which will be used in the simulations
at low THz intensities, we first fit to the splitting of the highest
THz intensity spectra to obtain the Rabi energy at this intensity.
Because the Rabi energy is proportional to the electric field, we
scaled the Rabi energies at lower powers by the square root of the
applied THz power. The method of scaling the Rabi energy by the square
root of the THz power is superior to fitting at every THz power because
it eliminates unnecessary fitting parameters and errors involved with
fitting peaks which are not fully separated. 

\section*{Appendix C. Heating}
\addcontentsline{toc}{section}{Appendix C. Heating}

Our observations cannot be the result of either lattice or carrier
heating. Though interband transitions are dependent on the energy
of the band gap and therefore are highly susceptible to temperature
changes, intersubband levels in the quantum well are determined by
the width of the well and the effective mass of the particle, neither
of which are strongly temperature dependent \citep{Cardona1961}. Hence,
any change in the temperature of the system should first be observed
as a shift of the exciton absorption line. This is shown in temperature
dependent measurements of the Autler-Townes effect performed by Sam
Carter (figure 3.19, page 80 of \citep{Carter2004}), where the Autler-Townes
peaks at 40K have the same relative amplitudes as the peaks at 10K,
but the main exciton line has been shifted by 2 meV. In the experiments
presented we observe no discernable shift of the exciton, only a large
red-shift of the excitonic $intersubband$ energy. We measure absorption
with a very weak LED ($<0.5\,\textrm{W/c\ensuremath{m^{2}}}$) and
are therefore insensitive to heating of the extremely small carrier
population. The negligible effects of heating are also consistent
with previous experiments performed by \v{C}erne \citep{Cerne1995}
and Carter \citep{Carter2005}.

\section*{Appendix D. Materials parameters}
\addcontentsline{toc}{section}{Appendix D. Materials parameters}

The following parameters were used in simulations of the InGaAs/AlGaAs
system described in this letter: a dielectric constant $\varepsilon=13$,
an electron effective mass $m_{\textrm{e}}^{*}=0.065\, m_{0}$, and
a hole effective mass $m_{\textrm{hh}}^{*}=0.48\, m_{0}$ for all
heavy holes.

\section*{References}{}

\addcontentsline{toc}{section}{Refrences}

\bibliographystyle{iopart-num}
\bibliography{BZaksBibNJP2}

\end{document}